# The flow of a very concentrated slurry in a parallel-plate device: influence of gravity

Marie Lenoble, Patrick Snabre and Bernard Pouligny
*Centre de recherche Paul-Pascal, CNRS, av. A. Schweitzer, 33600 Pessac, France*

ABSTRACT

We investigate, both experimentally and theoretically, the flow and structure of a slurry when sheared between two horizontal plates. The slurry, otherwise called "wet granular material", is made of non-Brownian particles immersed in a viscous fluid. The particles are *heavier* than the immersion fluid, in contrast to so-called "suspensions", corresponding to density-matched fluid and particles. Consequently, gravity influences the structure and flow profiles of the sheared material. Experiments are carried out in a plane Couette device with a model slurry composed of quasi-monodispersed spherical PMMA particles in oil, at high average solid concentration (about 59 %). Optical observation reveals a typical 2-phase configuration, with a fluidized layer in contact with the upper plate and on top of an amorphous solid phase. We provide data on velocity profiles, wall-slip, average shear stress and average normal stress, versus the angular velocity of the upper plate. To interpret the data, we propose a model for the ideal case of infinite horizontal flat plates (plane Couette flow). The model, of mean-field type, is based on local constitutive equations for the tangential ($\tau$) and normal ($N$) components of the stress tensor and on material expressions relating the material viscometric coefficients (the shear viscosity $\eta$ and the normal stress viscosity $\psi$) with the local concentration ($\Phi$) and the local shear rate. 1-, 2-, and 3-phase configurations are predicted, with non linear flow and concentration profiles. We conclude that model equations correctly describe the experimental data, provided that appropriate forms are chosen for the divergences of $\eta$ and $\psi$ near the packing concentration ($\Phi_m$), namely a $(\Phi_m - \Phi)^{-1}$ singularity.

Electronic mail:
lenoble@crpp-bordeaux.cnrs.fr,
snabre@crpp-bordeaux.cnrs.fr,
pouligny@crpp-bordeaux.cnrs.fr

# 1. INTRODUCTION

In the past two decades several studies have been devoted to the so-called "viscous re-suspension" phenomenon[1-4]. This term refers to the situation depicted in Fig.1: consider a sediment made of accumulated large (non Brownian) particles immersed inside a viscous fluid (Fig.1a). If the fluid layer on top is put in motion at moderate speed, a laminar shear flow results, which erodes the sediment. The particles from the eroded part form a fluidized suspension, below a pure fluid layer and on top of the remaining part of the sediment (Fig.1b). This configuration is a 3-phase system, with well defined boundaries, namely a sediment-suspension (SeSu) interface, and a suspension-pure fluid (SuF) interface. A well known model to interpret the phenomenon was worked out by Leighton and Acrivos[1]: essentially, the fluidized layer results from the balance between particle sedimentation and shear-induced particle diffusion. The model and observations were later refined[2], with a few contributions devoted to the stability of the SuF interface[5,6].

In the sequence illustrated in Fig.1 the fluid motion is forced by moving the upper horizontal plate in a plane Couette geometry. The above mentioned 3-phase equilibrium is observed at small plate velocity ($V$). At very high $V$, the system takes on a 1-phase configuration, as fluidization is complete and the suspended particles occupy the full space between the plates (Fig. 1e). What happens in the intermediate velocity range depends on the relative amounts of pure fluid and sediment in the initial state (Fig. 1a), i.e. on the average solid volume fraction ($\overline{\Phi}$). Let us start from the 3-phase state and increase $V$. In the case of a large fluid excess, the SeSu interface reaches the bottom plate first. The sediment disappears and what remains is a 2-phase state, made of a suspension and of a pure fluid layer (Fig. 1c). Conversely, in the case of a small fluid excess, the SuF interface reaches the top plate first, resulting in a sediment-suspension 2-phase state (Fig. 1d).

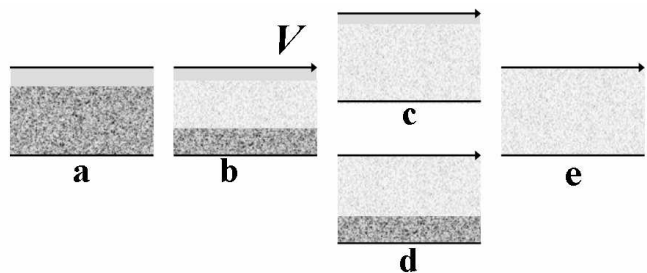

Fig. 1: Scheme of the viscous re-suspension problem. The top plate of the shear device is moving, as indicated by the arrow, from left to right. The system under shear adopts a 3-phase (b), 2-phase (c or d) and 1-phase (e) configuration with increasing plate velocity.

This paper addresses the latter situation, and is mainly devoted to the 2-phase regime. Our samples are very concentrated, i.e. the "sediment", corresponding to the initial situation in Fig.1a almost completely fills the gap between the two plates. In the real experimental procedure, the upper plate is initially brought in contact with the sediment. Because the sample is basically a granular material, particles in contact to one another form chains that prevent complete compaction under the sole action of gravity. The first action of the shear is to break the chains and to operate compaction. Afterwards the system takes on a stationary 2-phase geometry, with only the upper part of the system being sheared. We provide experimental observations in the form of flow profiles, shear stress and normal stress measurements in this regime and set up a model to interpret the data. A central goal is to account for the suspension thickness as a function of the plate velocity. We will consider the suspension as a continuum, characterized by simple viscometric coefficients, namely shear ($\eta$) and normal stress ($\psi$) viscosities. As the average particle concentration ($\overline{\Phi} \cong 59\%$) is not far from the random close packing, $\Phi_m \cong 64\%$, the response of the system is very sensitive to the functional forms of $\eta$ and $\psi$ near $\Phi_m$. In this respect, simply observing the 2-phase structure bears information about this point.

The paper is structured as follows:

- Section 2 is about the experimental investigations and procedures. Briefly, we study a model slurry made of polymethacrylate spheres immersed in an index-matched oil. Samples are sheared in a plane Couette (parallel-disk) device, and observed for varying plate angular velocity ($\Omega$). As the whole system is transparent, it is easy to view the 2-phase structure using particle image velocimetry. Flow profiles are obtained using fluorescent tracers, and video image analysis. The mechanical response of the sheared sample is characterized through the torque and the normal axial force acting on the rotating plate, at imposed rotor speed or imposed shear stress.

- Observations and data are gathered in Section 3.

- The model is worked out in Section 4. We follow the principles of the "suspension balance theory"[7-9], by setting out constitutive equations for both the shear and normal components of the material stress tensor, and adopt simple generic forms for $\eta$ and $\psi$ as functions of the local particle concentration, $\Phi$. The equations for the concentration and flow profiles are solved numerically; but the main features of the solutions are obtained through simple graphic constructions and asymptotic behaviors.

- In Section 5, we compare the predictions of the model to experimental data, with a focus set on the functional forms of $\eta$ and $\psi$ close to compaction. As the experiments show that the suspended layer partially slips along the upper plate, it is important to make a distinction between the plate velocity $V$, and the velocity of the particle layer closest to the plate, $V_0$. As we will see, this point about wall-slip is essential when comparing the model to the data.

- The paper is summarized and concluded in Section 6.

## 2. EXPERIMENTAL PROCEDURE
*2.1 Sample*

The model slurry is constituted of polymethylmetacrylate (PMMA) spherical particles immersed in a hydrophobic fluid. The particles (density 1.19 g.cm$^{-3}$) are heavier than the fluid and are sieved to diameters between 180 and 200 micrometers. The fluid, a mixture of hexadecane and of microscope oil (Sigma-Aldrich S150), has a density $\cong 0.89$ g.cm$^{-3}$, and a shear viscosity $\eta_o \cong 25$ cp at room temperature. As the corresponding solid-fluid density mismatch ($\Delta \rho \cong 0.3$ g.cm$^{-3}$) is large, the sample sedimentation in the shear cell takes place within a few minutes.

The samples, about 4 cm in the largest dimension, are about transparent at room temperature. They show a slight opalescence due to imperfect index-matching. To characterize the flow, we use markers, i.e. PMMA particles from the same batch which are made fluorescent by impregnation of an organic dye (Rhodamine 6G). The impregnation is performed by immersion of the particles in an ethanol solution of the dye, for about half an hour at 40°C, followed by rinsing in water and drying. Microscope observation shows that the dye slightly penetrates the particle and remains trapped in a peripheral skin, about a micrometer in thickness. When the tracers are immersed in the hydrophobic immersion fluid, they show an intense orange fluorescence under green illumination (Argon ion laser, 514 nm line), with no detectable leak of the dye out of the particles.

*2.2 Shear apparatus* (Fig.2)

The device is basically the bottom part of a Couette cell : the top plate is the bottom flat surface of a cylinder (radius $R_1$=18.5 mm) that penetrates into a coaxial cylindrical cavity (radius $R_2$=20 mm), with a flat bottom too. The distance between both horizontal plates, namely the gap $d$, usually spans a few millimeters. Initially the slurry is prepared inside the cavity. The volume of oil is larger than the minimum amount necessary to fill the spaces between the particles; therefore the system is prepared as a sedimentary bed of particles completely immersed in the fluid (Fig.2a). Afterwards the inner cylinder is brought down to contact with the sediment, and pushed a little bit lower. As a result, a small part of the sediment is expelled out and pushed up in the annular region between both cylinders (Fig.2b). This procedure allows us to prepare the sample at very high solid volume fraction, $\overline{\Phi} \cong 59$ %. This value is located between the so-called "random loose packing" ($\cong 56$ %) and "random close packing" ($\cong 64$ %) volume fractions, which means that the material initially is supported by chains of particles[10]. Afterwards, when shear is in operation, these chains are partially broken, resulting in further compaction of the material. This shear-induced compaction is visible in the ultimate stage of the experiment when the machine is stopped and the sample left at rest: the particles are not in contact with the top plate any more, leaving a layer of pure oil on top of the sample, as sketched in Fig. 1a. But the thickness of this layer (denoted $\delta$ below) is very small, on the order of 2 particle diameters.

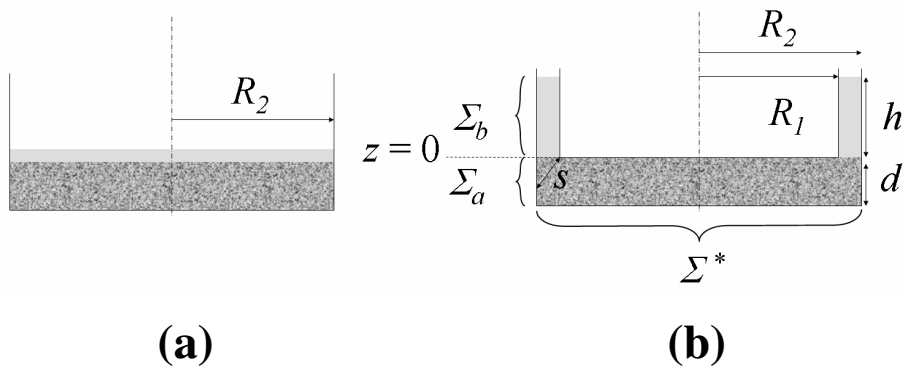

**(a)** **(b)**

Fig. 2: Couette cell and shear experiment preparation. a) Initially the slurry is prepared with a slight excess of immersion fluid inside the stator (radius $R_2$) . b) The rotor (radius $R_1$) is brought in contact with the sediment. The excess immersion fluid forms a ring of height $h$ between the coaxial cylinders. See Appendix A for definitions of other symbols.

The whole device is made of PMMA, to make the experimental cell fully transparent. Both cylinders are driven in rotation by independent motors[11]. In most experiments, only the inner cylinder is moving, and is called "rotor" in this context. As noted in earlier studies[12], granular materials show considerable -often complete- slip on smooth surfaces. For this reason, the surfaces of the Couette cell in contact to the sample are made rough on the scale of the particle size, using a simple wood file. The procedure gives random scratches, a few 100 µm in width and a few 10 µm in depth. Wall-slip is not eliminated, as we will see, but greatly reduced.

*2.3 Optics* (Fig.3)

In our setup, fluorescence is excited by means of a vertical laser sheet (LS, wavelength: 514 nm) that can be positioned at variable distances from the cell axis ($0 \leq r \leq 20$ mm), and is observed through an orange filter (F). Images show the repartition of markers in vertical planes. In an alternate version, the laser sheet is set horizontal, and the optical observation is performed from below through the bottom of the Couette cell, giving a map of the flow at constant altitude ($z$) and variable $r$.

Flow profiles are obtained by analyzing sequences of digital video frames from the CCD camera. A homemade "Digital Particle Image Velocimetry" (DPIV)[13] software was developed to deduce the average displacement field of tracers from the cross correlation of corresponding subareas in successive frames. The location of the highest correlation peak corresponds to the most likely particle displacement in the interrogation subarea. Sub-pixel accuracy is obtained by Gaussian interpolation of the correlation peak in the $x$ and $y$-directions. If the displacement vector exceeds a prescribed amount (on the order of the size of the interrogation area), the vector is considered as spurious and removed. The DPIV accuracy was tested using artificially generated image sequences of uniform displacement and solid body rotation.

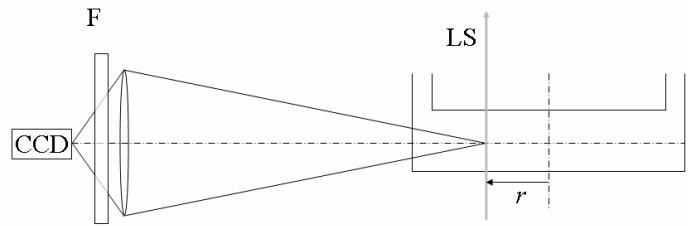

Fig. 3: Scheme of the optical setup. In this configuration, the experiment provides a view of fluorescent tracers inside a vertical plane across the Couette device.

When the laser sheet is set vertical, the method provides a direct view of the flow at the distance $r$ from the axis and at $\theta = \pi/2$ (taking the origin of the polar angle on the optical axis towards the camera), as a function of altitude. In typical conditions, a DPIV sequence comprises $N \approx 100$ video frames, and the frame rate is chosen proportional to the average velocity of particles in contact to the rotor. The method provides $N$-1 snapshots of the displacement field, from which an average velocity field is deduced.

In principle, *concentration fields* can be obtained from the map of fluorescence intensity through a two-fold assumption: i) the fluorescence intensity is everywhere proportional to the tracer concentration ($\Phi_{tr}$); ii) the latter is itself proportional to the concentration of bare particles, $\Phi$. Condition i) can be satisfied if $\Phi_{tr}$ is made small enough (say, not more than 3 % of the particles are tracers), but condition ii) turns out much more challenging. In fact, we found that minute differences in the size distribution of tracers compared to that of the bare particles were enough to provoke visible changes in the intensity profiles. The phenomenon is due to a slight segregation of the tracers. The segregation is not as intense as that observed with strongly polydisperse systems[11], but with the same tendency (smaller particles go to the bottom, near the SeSu interface). Note that another complication may come from the fact that the surfaces of the fluorescent tracers are different from those of the bare particles, and then a surface-driven partial particle segregation[14] cannot be ruled out a priori. Because of these difficulties, and others related to the collection efficiency of the optical imaging system, we did not attempt to elaborate quantitative concentration profiles, and left this as a goal of a future work.

*2.4 Rheometry*

Rheology experiments were performed at room temperature (25°C) with a TA Instrument AR2000 rheometer in a homemade transparent PMMA Couette cell similar to that used for flow characterization, consisting of a rotating inner cylinder ($R_1$ =18.5 mm) in a fixed outer cylinder (radius $R_2$ =20 mm). The instrument is equipped with a force transducer to measure the axial force $F_N$ acting upon the rotor. The procedure to prepare the sample and to bring the rotor in contact to the particles is the same as that described in Section 1, giving a mean particle volume fraction $\bar{\Phi} \cong 59$ %. The height ($h$) of excess fluid in the annular volume between the cylinders is about 10 mm. The rheological behavior of the slurry is investigated in the controlled shear mode by a stepwise increase of the angular velocity $\Omega$ of the inner cylinder and record of both the viscous torque $\Gamma$ and the total axial force $F_N$ acting upon the inner cylinder. The time duration of the angular velocity ramp was set = 4 hours, which turned out slow enough to avoid hysteresis in $\Gamma$ and $F_N$ when the ramp was reversed in time. This indicates that sufficient time was allowed for relaxation of the slurry microstructure at every step within the experiment time. The instrument yields an average shear stress $\bar{\tau} = \Gamma /(\pi R_1^3 / 2)$ and an average normal stress $\bar{\tau}_N = F_N / (\pi R_1^2 /2)$. The sensitivity of the axial force sensor is about 5 10$^{-3}$ N, so that the resolution in $\bar{\tau}_N$ is about 10 Pa.

In Section 5, we analyze the experimental rheological response on the basis of the model worked out below in Section 4 for a simple shear geometry, i.e. between two horizontal infinite planes. The comparison is complicated by the finite size of the experimental Couette cell, which imposes a vanishing velocity at $r = R_2$ (supposing a no-slip condition), and by the presence of the viscous fluid in the annular space between both cylinders. Both factors contribute to increasing the measured torque beyond the "ideal" value, $\bar{\tau}*$. To deduce $\bar{\tau}*$ from $\bar{\tau}$, we empirically worked out an approximate correction formula, which we tested with the pure index-matched fluid in the Couette cell. Details of the procedure and the correction formula (Eq. A3) are given in Appendix A. Note that this correction only applies to the torque ($\bar{\tau}$), not to the axial force ($\bar{\tau}_N$). Working out a correction formula for $\bar{\tau}_N$ would need replacing the sample by a nonlinear fluid of known normal force characteristics, a more ambitious task that we do not perform yet. In the following (Section 3), we only give bare (uncorrected) data for $\bar{\tau}_N$.

## 3. RESULTS
*3.1 Flow*

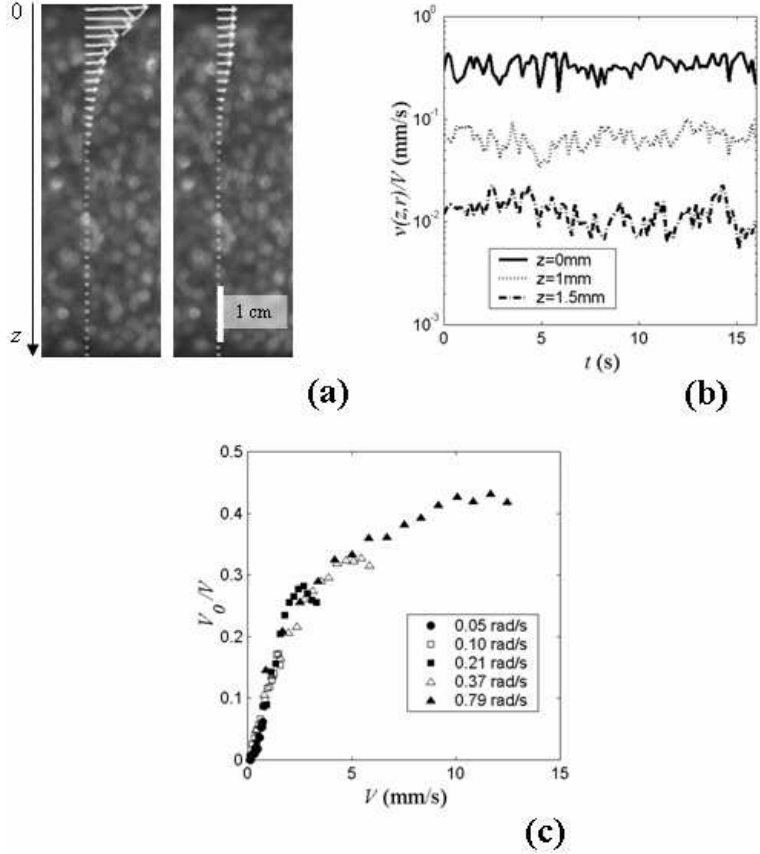

Fig. 4: Fluctuations of the flow and wall-slip. a) The photos are snapshots of the velocity field in a vertical plane. Note the large fluctuations of the DPIV vectors in the flow direction. b) Time fluctuations of the horizontal component of the velocity, at different depths, scaled to the plate velocity for $r =12$mm and $\Omega = 0.262$ rad/s. c) The coupling of the suspension to the upper plate fluctuates in time. The wall-slip effect is defined through the $V_o/V$ ratio, where $V_o =< v(z = 0, r) >$. The coupling is very poor at small speed, but improves up to a plateau value in the high speed regime.

The sequence of PIV snapshots in general reveals time fluctuations in the flow velocity, especially in the flow direction (see Fig.4a, b). For particles near the rotating plate, the mean quadratic fluctuation velocity in the flow direction is about 10% to 15% of the local plate velocity, $V = \Omega r$, while velocity fluctuations in the perpendicular direction are an order-of-magnitude lower. A second - but apparently related – feature is the fact that the average velocity of the particle layer in contact to the upper plate, $V_o = v(z = 0, r)$, is systematically smaller than the plate velocity $V$, i.e. the suspension partially slips along the plate. This point is illustrated by the graph in Fig. 4c, showing the variation of $V_o/V$ versus $V$. This graph gathers data obtained either at constant $\Omega$ and variable $r$, or at constant $r$ and variable $\Omega$. Note that the points approximately merge onto a single curve, indicating that $V_o/V$ only depends on the local plate velocity $V = \Omega r$. Clearly, wall-slip is almost total at small $V$, while the coupling of the suspension to the plate is improved at high rotor speed, but is never better than about 50 %.

Examples of velocity profiles, $v(z,r)$, are shown in Fig. 4a, as a function of altitude at constant $r$. The $z$ axis is directed downwards, with the origin taken on the top plate (therefore the bottom plate is at $z = d$). $V$ is taken as the unit velocity in the plots. The figure shows how the flow profile depends on the local plate velocity, $V = \Omega\, r$. All profiles have about the same non linear shape. The local shear rate, i.e. the slope of the velocity profile, is highest near the upper plate. Each profile falls down to zero within a distance $\zeta < d$, i.e. the structure is of 2-phase type (Fig.1d). The length $\zeta$ is defined as the thickness of the sheared region at the distance $r$ from the axis.

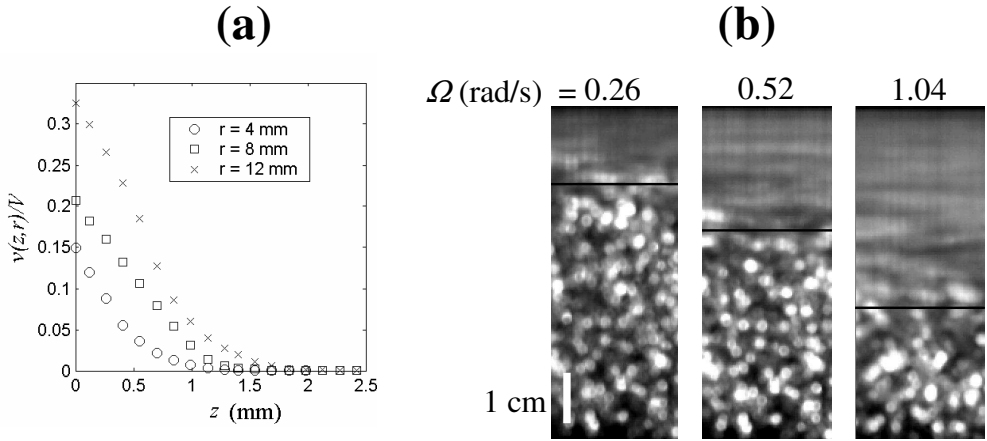

Fig. 5: Average shear flow. a) Velocity profiles, scaled to $V$, at constant $\Omega$ and variable $r$. b) Averaged video sequences at variable $\Omega$ and constant $r$ (= 12mm in this example) used to determine the thickness $\zeta$ of the flowing region.

The local 2-phase structure is most simply revealed by averaging the frames of the DPIV sequence, as shown in Fig. 5b: immobile particles are distinctly visible in the lower part of the image ($z > \zeta$), while the upper part ($z < \zeta$) appears blurred. In principle, the value of $\zeta$ is simply the thickness of the blurred zone. Experimentally, we define the limit of the blurred zone as the altitude $z_c$ where the particles move by about one diameter ($2a$) within the duration of the PIV sequence, $NT$. This definition is based on the representation of the velocity field as $v(z,r) = V_o\, f(z/\zeta)$, which we justify in the following Sections 4 and 5. The critical altitude $z_c$ is simply given by $z_c = \zeta\, f^{-1}(2a/V_o NT)$. Since $NT$ is inversely proportional to $V_o$, $2a/V_o NT$ is a constant $\ll 1$, and $f^{-1}(2a/V_o NT)$ is close to unity. Consequently, the experimentally measured $z_c$ is proportional to and close to $\zeta$.

The variation of $\zeta$ (i.e. $z_c$) with the plate velocity is shown in Fig.6a. The figure is a collection of three $\zeta(r)$ curves, corresponding to three different rotor speeds ($\Omega =$ 0.26, 0.52 and 1.05 rd/s, from left to right). In each case, $\zeta$ first increases with $r$ (up-branch), goes through a maximum and decreases (down-branch). As each data ends at $r \cong 18$ mm, i.e. close to the outer wall of the Couette cell, it is clear that the down-branches are due to the finite extent of the experimental device: the velocity must vanish along the stator. Interestingly, the up-branches approximately merge on a single curve, which may be viewed as the ideal $\zeta(V)$ curve expected for a shear device of infinite diameter.

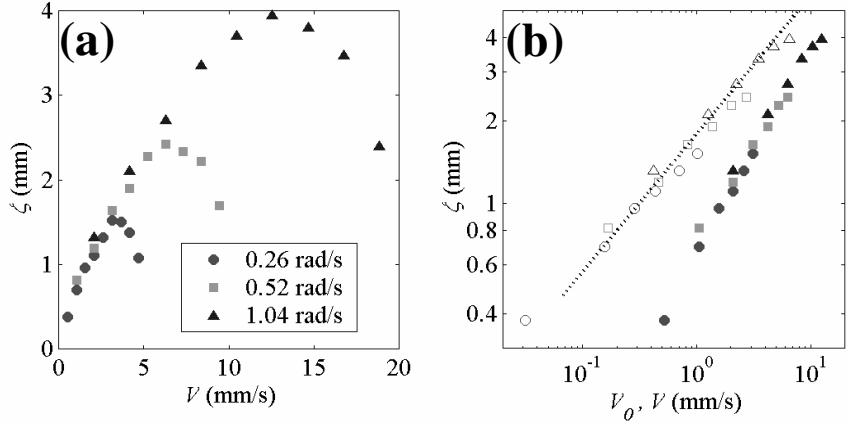

Fig. 6: Thickness $\zeta$ of the flowing region versus local plate velocity $V = \Omega r$. a) The graph is composed of three $\zeta(r)$ curves corresponding to three rotors speeds. Finite-size effects due to coupling to the outer boundary of the shear device are responsible for the decrease of $\zeta$ at the cell periphery. b) $\zeta$-versus-$V$ (filled symbols) and $\zeta$-versus-$V_o$ (open symbols), in log-log representation. The dotted straight line is the prediction of the model in the asymptotic regime (Eq.(19)), with $\alpha = 1$, $\omega = 0$, $p = 5.5$, $\eta_o = 0.25$ Poise, $\psi_o = 9.2$ Poise, $\lambda = 188$ g.cm$^{-2}$.s$^{-2}$.

Since $V_o$ rather than $V$ is the reference velocity of the flow profile, an alternate and instructive representation is the $\zeta$-versus $V_o$ graph (open symbols, Fig. 6b). In Fig.6b, data from Figs.6a and 4c are gathered, now in log-log representation. Filled symbols are used for the $\zeta$-versus $V$ graph. For clarity, the points corresponding to large radii (finite-size effect) have been removed. Note that the dynamical range of $V_o$ (2 decades) is larger than that of $V$, because of the considerable wall-slip at small velocity, and that the "ideal" $\zeta(V_o)$ curve is approximately linear in log-log scale, suggesting a power-law relationship: $\zeta \propto V_o^{\varepsilon}$, with $\varepsilon \cong 0.46$.

*3.2. Rheometry*

Experiments show similar rheological behaviors for the slurry whatever the angular velocity ($\Omega$) or the average torque ($\Gamma$) is controlled (Fig.7a-c). The normal force is directed upwards, which means that the sheared sample "wants" to increase the gap between the plates, as one might expect because of the dilatancy effect. In the range $0 \leq \Omega \leq 2$ rad/s, the rheological response of the slurry is reversible after stepwise increase and subsequent decrease of the angular velocity or torque. Hysteresis only shows up in the high shear regime ($\Omega > 2$ rad/s), apparently due to an irreversible migration of particles toward the thin annular region between cylinders.

The corrected average shear stress ($\bar{\tau}^*$) and the average normal stress $\bar{\tau}_N$ show similar continuous variations in $\Omega$. Indeed, both responses are found proportional in the full experimental range (Fig.7c), within experimental noise. At low rotor speed, $\Omega < 0.3$ rad/s, $\bar{\tau}^*$ and $\bar{\tau}_N$ are about second order in $\Omega$. This vanishingly small rheological response is related to the considerable wall-slip in this velocity range ($0 \leq V = \Omega r \leq 5$ mm/s), which ruins the efficiency of the rotor in shearing the bulk of the sample. At higher rotor speed, $\Omega > 0.5$ rad/s, wall-slip tends to saturate (see Fig.4c), and both $\bar{\tau}^*$ and $\bar{\tau}_N$ linearly increase with $\Omega$ (Fig.7a, b). Note that there is no indication of a yield shear stress from the $\bar{\tau}^*(\Omega)$ curve (Fig.7a).

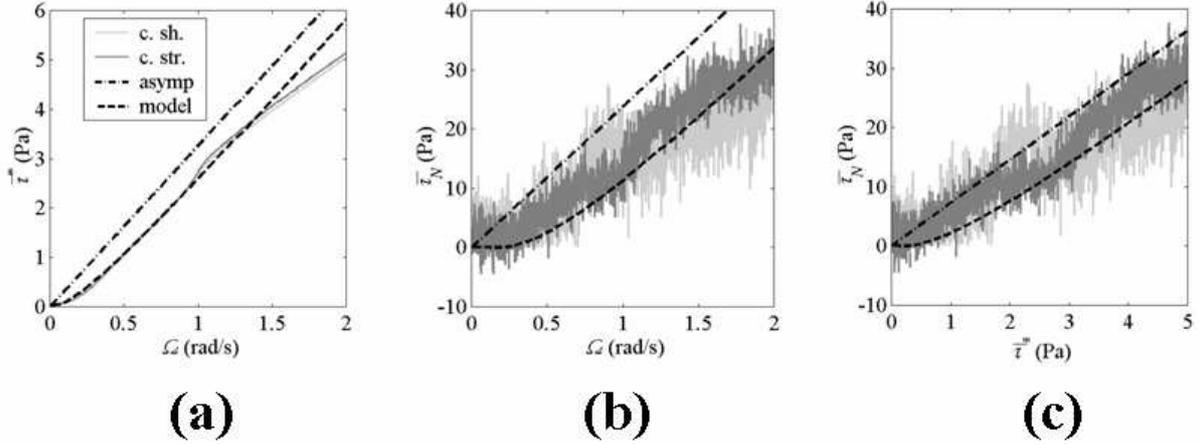

Fig. 7: Average shear stress $\bar{\tau}^*$ (*a*) and average normal stress $\bar{\tau}_N$ (*b*) versus angular velocity of the rotor ($\Omega$) in controlled shear mode (c.sh.) or controlled stress mode (c.str.) for mean particle volume fraction $\overline{\Phi} = 0.59$ and gap $d = 5$mm. c : average normal stress $\bar{\tau}_N$ versus average shear stress $\bar{\tau}^*$. The dashed line is the prediction of the model, according to Eq.(22), with $\omega = 0$, $p = 5.5$, $\eta_o = 0.25$ Poise, $\psi_o = 9.2$ Poise, $\delta = 390$ μm, $R_1 = 18.5$mm, $\mu \approx 0.6 \Omega r /(V_c + \Omega r)$ and $V_c = 4$mm/s. The straight line (mixed) corresponds to the asymptotic relation Eq.(23) with $\mu_\infty = 0.6$.

Complete re-suspension, i.e. the transition from the 2-phase to 1-phase configuration occurs at $\Omega \approx 1$ rad/s near the periphery of the rotor. This transition is not signaled in the rheological response; in other words, rheometry does not provide a direct information about the flow localization inside the sheared material.

## 4. MODEL
*4.1 Basics*

We consider the ideal shear geometry sketched in Fig.8: the sample is sheared between two infinite horizontal planes, separated by a distance $d$. The bottom plate is fixed, while the upper one moves horizontally, with an imposed velocity, $V$ in *x*-direction. The depth inside the sample, starting from the upper plate, is denoted $z$.

The goal of the model is to derive functional forms for the *average* concentration and velocity fields, respectively. In principle, a complete theory should be a kind of statistical mechanics of the shear flow, able to predict both the average and the "noise" of the relevant fields[15]. In fact, experimental observations show that the noise in the velocity field is everywhere considerable, even in regions that do not flow in average. Such a complete theory is definitely beyond the scope of this article, as we will content ourselves with predicting average concentration and velocity fields, namely $<\Phi(x,z)>$ and $<v(x,z)>$. Averaging leaves a simple $z$ dependence; consequently both fields will be simply denoted $\Phi(z)$ and $v(z)$. The model is of "mean-field" type, as it amounts to setting up equations that only couple average fields.

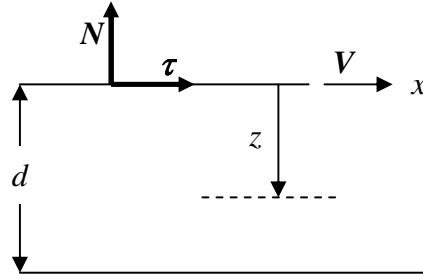

Fig. 8: Scheme of the theoretical infinite-plate geometry. $\tau$ and $N$ are the tangential and normal stresses exerted by the sheared slurry upon the upper plate.

As illustrated in Fig.1, the shear lifts particles from the sediment to form the fluidized phase (Su). The fluidization may be viewed as due to a shear-induced normal stress, proportional to the local shear rate :

$$\mathbf{n} = -\psi \, \dot{v} \, \hat{z} \qquad (1)$$

where $\dot{v} = -dv/dz$ is the local shear rate, $\hat{z}$ is the unit vector along $z$, and $\psi$ the « normal stress viscosity »[8]. If $\mathbf{n}$ were an analytical function of the shear rate, symmetry would impose that $\mathbf{n}$ be proportional to the square of $\dot{v}$, to lowest order in $\dot{v}$. This scaling is verified with visco-elastic polymer solutions[16], but apparently not by concentrated non-colloidal suspensions. In this case, experiments[17] and simulations[18,19] indicate that the normal components of the stress tensor scale as the absolute value of $\dot{v}$. Hence the linear form in Eq. (1).

In the 2-phase and 1-phase regimes (Fig. 1d, e), the suspension is in contact with the upper plate, creating an upward stress ($-N\hat{z}$) against the plate. A reaction stress, $N\hat{z}$, is opposed by the plate to keep the gap $= d$. Note that in the 3-phase regime (Fig. 1b), the upper plate is in contact with a layer of the pure Newtonian fluid; therefore $N = 0$ in this case.

The viscous normal stress $\mathbf{n}$ is opposed by the sum of $N$ and of the weight per unit area of the suspended layer between 0 and $z$, corrected for buoyancy. The balance equation reads:

$$\psi \dot{v} = N + \lambda \int_0^z \varphi \, dz' \qquad (2)$$

Here $\varphi = \Phi/\Phi_m$ is the reduced concentration, $\Phi_m$ the particle concentration in the sediment. $\lambda$ is the gravity parameter, defined as $\lambda = \Delta\rho \, g \, \Phi_m$, where $g$ is the gravity acceleration and $\Delta\rho = \rho_S - \rho_F$ is the density difference between solid and fluid.

In the horizontal direction, a stress $\tau \hat{x}$ is applied to the upper plate. The tangential stress $\tau \hat{x}$ is opposed to the viscous stress, which, for a Newtonian fluid, is simply proportional to the shear rate, i.e. $\tau \propto \eta \dot{v}$, where $\eta$ is the suspension shear viscosity. A very concentrated suspension departs from this simple rule at least because gravity jams the material. Accounting for jamming effects in the flow of granular matter is a very complex matter in general[20-22]. Here we will simply model the influence of $g$ on $\tau$ as a $z$-dependent threshold, $\tau_S$. To justify this point, we consider a particle at depth $z$ in the initial sediment (Fig. 1a), and suppose that a horizontal force is exerted on the particle to put it in motion. Because the system is jammed by gravity, the particle cannot be moved on a macroscopic distance unless the force is greater than a minimum, which is on the order of the weight of the particles column above $z$. We suppose that such a picture remains valid in steady state motion, and take into account both the $N$ contribution and the weight per unit area of the particle column in the expression of the threshold :

$$\tau_S = \omega( N + \lambda \int_0^z \varphi \, dz' ) \qquad (3a)$$

$$\tau = \tau_S + \eta \dot{v} \qquad (3b)$$

In Eq. (3a), $\omega$ is a proportionality constant that is of order 1 according to the above argument, or that can be put $= 0$ if the threshold effect is ignored or turns out irrelevant. As we will see in the next section, analysis of the experimental data allows us to decide on this point.

Eqs. (2) and (3) are equations of motion for the concentration and velocity fields, $\varphi(z)$ and $v(z)$, respectively. The first boundary condition is mass conservation, i.e.

$$\int_0^d \varphi \, dz = d \, \overline{\varphi} \quad , \qquad (4)$$

where $\overline{\varphi}$ is the average reduced concentration. Boundary conditions for the velocity field are $v(z=d)=0$ and $v(z=0)=V_o$. Recall that $V_o < V$ because of the wall-slip effect along the upper plate. We will not attempt to find a theoretical relation between $V_o$ and $V$, because it would need elaborating a model for the suspension wall-slip. At the current stage, it is unclear to us whether this model should be closer to the Coulomb theory of solid friction or to the wall-slip effect of molecular fluids[23-25].

The viscometric coefficients in Eqs.(2, 3) only depend on the particles local concentration. General functional forms are:

$$\eta = \eta_o \, p(\varphi) \, (1-\varphi)^{-\alpha} \qquad (5)$$

$$\psi = \psi_o \, q(\varphi) \varphi^\beta (1-\varphi)^{-\alpha'} \qquad (6)$$

$\eta_0$ is the viscosity of the interstitial fluid. The viscometric coefficients are supposed to diverge near compaction $(\varphi = 1)$ according to power-laws with exponents $\alpha$ and $\alpha'$. In the limit of infinite dilution $(\varphi \to 0)$, the suspension viscosity $\to \eta_o$, and since the pure fluid is Newtonian, $\psi \to 0$; hence the $\varphi^\beta$ factor in Eq.(6). $p(\varphi)$ and $q(\varphi)$ are regular ("smooth") functions, that neither go to zero nor diverge in the [0,1] interval.

Different explicit forms for $\eta$ have been proposed in the literature, with $\alpha = 1$ (Frankel and Acrivos[26], Nunan and Keller[27]), $\alpha = 2$ (Brady[28, 29], Mills and Snabre[30], Morris and Boulay[8]), $\alpha = 3$ (Zarraga et al.[9]). $\alpha = 1.82$ in Krieger's model[31]. Literature data about the normal force coefficient is scarce, again with different possibilities for the values of the exponents: $\beta = 2$, $\alpha' = 2$ (Morris and Boulay[8]), $\beta = 3$, $\alpha' = 3$ (Zarraga et al.[9]). In the following, we will just suppose that $\alpha$, $\beta$ and $\alpha'$ are positive integers, and that $\alpha = \alpha'$ (i.e. that $\eta$ and $\psi$ diverge in the same way near compaction). The latter assumption, which has the advantage of greatly simplifying the resolution of the equations, is mainly supported by our finding that the rheological responses $\overline{\tau}*$ and $\overline{\tau}_N$ are proportional (Fig.7c). As we will see in the next section, data analysis confirms this view.

*4.2 Concentration profile and phase sequence.*

We define a characteristic length, $l = \tau/\lambda$, and a reduced depth, $\bar{z} = z/l$. Eliminating $\dot{v}$ between Eq.(2) and Eq.(3) yields the equation for the concentration profile:

$$\frac{d\bar{z}}{d\varphi} = \chi_\omega(\varphi) \qquad , \qquad (7)$$

with

$$\chi_\omega(\varphi) = \frac{1}{\varphi}\frac{d}{d\varphi}\left[\frac{1}{\omega + (\eta/\psi)}\right] \qquad (8)$$

In the $\varphi \to 0$ limit, $d\bar{z}/d\varphi$ behaves as $\varphi^{\beta-2}$. This limit is encountered in the 3-phase regime (Fig. 1b), and it is known from the observations by Leighton and Acrivos[1] that the interface between the suspension and the pure fluid is sharp. Consequently, $d\bar{z}/d\varphi$ must vanish near $\varphi = 0$, imposing $\beta > 2$. We now put $\beta = 3$, which meets the above condition and is in agreement with the form proposed by Zarraga *et al.*[9].

Integrating Eq. (8) yields the concentration profile, $\bar{z}(\varphi) = X_\omega(\varphi) + \bar{z}_1$, where

$$X_\omega(\varphi) = -\int_\varphi^1 \chi_\omega(\varphi')d\varphi' \qquad (9)$$

$\bar{z}_1$, the depth corresponding to $\varphi = 1$, is an integration constant such that the mass conservation, Eq.(4), be satisfied. The general shape of the profile is shown in Fig.9a. Different choices for $p(\varphi)$ and $q(\varphi)$ functions (these enter the $(\eta/\psi)$ ratio in Eq. (8)) yield profiles that of course are quantitatively different, but qualitatively display the same shape, with an inflexion point for a concentration denoted $\varphi_1$ in Fig. 9a. $\varphi_1$ may be either $<1$ or $>1$, depending on whether $\omega$ is large or small compared to $\eta_o/\psi_o$, respectively. If the jamming action of gravity is negligible ($\omega \to 0$), the inflexion is outside of the physical concentration range ($0 \leq \varphi \leq 1$), and the concentration profile is convex. Conversely, if $\omega$ is large, the profile has a characteristic S-shape, and is concave near the compaction limit ($\varphi = 1$).

We now turn to the sequence of configurations that appear when the stress, or the upper plate velocity, is increased, from very small to high values. We consider the initial situation sketched in Fig.9b: the sediment ($\varphi = 1$) fills most of the space between the plates, leaving a shallow layer, of thickness $\delta \ll d$, of pure fluid in contact to the upper plate. Consequently, the average concentration, $\bar{\varphi}$, is only slightly inferior to 1. The sequence of Figs.9c-e shows $z(\varphi)$ profiles for increasing values of $\tau$. Each profile is deduced from the universal $\bar{z}(\varphi)$ curve through a dilation of the ordinate axis by a factor $l = \tau/\lambda$, proportional to the shear stress $\tau$. In the low $\tau$ regime, one observes a 3-phase (Sediment-Suspension-Pure fluid) repartition, as in the model of "viscous resuspension"[1]. The thickness $\zeta$ of the suspension is simply given by $\zeta = l X_\omega(0)$. When $\tau$ increases, the top of the suspended layer hits the upper plate, resulting in a 2-phase configuration (Fig.9d). The vertical position of the $z(\varphi)$ profile is adjusted to satisfy mass conservation, which amounts to equating the grey areas in Figs.9b and 9d. Now $\zeta = l X_\omega(\varphi_o)$, denoting $\varphi_o$ the particle concentration at the upper plate. An even larger $\tau$ results in the total fluidization of the sample (Fig.9e). The dilation and position of the $z(\varphi)$ profile are obtained through the same graphical construction as in Fig.9d.

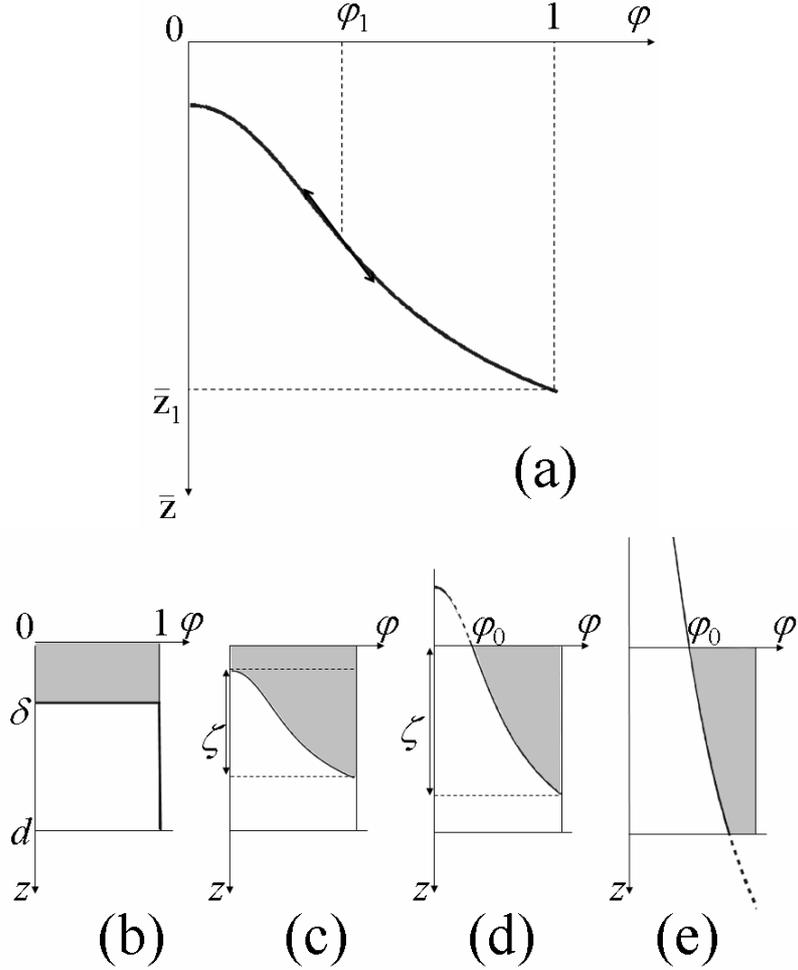

Fig. 9: Graphical construction to derive concentration fields. a) Shape of the master $X_\omega(\varphi)$ curve (Eq.(9)). The inflexion point is inside the $[0,1]$ physical concentration range only if $\omega$ is large. b) The sample initial state, in theory, is a 2-phase configuration, with a small layer (thickness $\delta$) of immersion fluid on top of a sediment ($\varphi = 1$). c-e) Sequence of 3-, 2-, and 1- phase configurations corresponding to increasing values of $\tau$.

Here we have discussed the case of a very concentrated sample. We end this paragraph by a remark about the classical viscous re-suspension problem. The typical initial situation in this problem involves a sedimentary bed immersed in a thick layer of fluid, and, compared to the situation sketched in Fig.9b, corresponds to the alternate condition $d - \delta \ll d$, with $\bar{\varphi} \ll 1$, correspondingly. The same kind of reasoning applies to this case, again leading to a sequence of 3-, 2- and 1-phase layering. But in this $\bar{\varphi} \ll 1$ regime, the 2-phase configuration involves a suspension and a pure fluid layer, instead of the sediment-suspension couple in the $\bar{\varphi} \cong 1$ regime.

*4.3 Velocity profile and rheological equations*
Eq.(2) yields:

$$\dot{v} = \frac{1}{\psi(\varphi)} \left[ N + \lambda \int_0^z \varphi \, dz' \right] \qquad (10)$$

and, putting $z = 0$ in Eq.(2):

$$\tau = N\left[\omega + \frac{\eta(\varphi_o)}{\psi(\varphi_o)}\right] \qquad (11)$$

Injecting the known concentration profile, $\varphi(z)$, into the above equations leads to a differential equation for the velocity profile, with $z$ as the variable. An alternate way is to build the $v(\varphi)$ function, which is solution of the following equation:

$$\frac{dv}{d\varphi} = -\tau \frac{l\chi_\omega(\varphi)}{\psi(\varphi)}\left[\frac{1}{\omega + \eta(\varphi_o)/\psi(\varphi_o)} + \int_{\varphi_o}^{\varphi} \varphi' \chi_\omega(\varphi')\, d\varphi'\right] \qquad (12)$$

Solving Eq.(12) yields the relation between velocity and concentration, with $\tau$ as a parameter, namely a $v(\varphi,\tau)$ function, and then the velocity profile, $v[\varphi(z),\tau]$. Afterwards, writing $V_o = v(\varphi_o,\tau)$ yields the relation between $\tau$ and $V_o$. We thus arrive at an expression for the velocity profile with $V_o$ as a control parameter, and can directly compare this expression to the experimental profile.

We built a numerical routine that follows this procedure and whose main results are given in Appendix B. Interestingly, an approximate analytical solution can be worked out in the $\delta \ll d$ limit (Fig.9b), which corresponds to our experiments. In this limit, the grey surface area in Fig.9b-d is very small, and the part of the $X_\omega(\varphi)$ curve which is represented by a solid line in Figs.9d-e can be considered as linear. This approximation is valid whenever $\tau$ is not too small, i.e. when $\zeta$ is definitely larger than $\delta$. In this limit and for the 2-phase regime, expanding the equations to leading order in $(1-\varphi)$ and using the graphical trick based on equating the grey areas in Fig.9 yields the following asymptotic forms:

$$\zeta = l\, X_\omega(\varphi_o) \cong l\,(1-\varphi_o)\,\chi_\omega(1) \qquad (13)$$

and

$$1 - \varphi_o \cong \sqrt{2\delta / l\chi_\omega(1)} \cong \frac{2\delta}{\zeta} \qquad (14)$$

The concentration profile is simply linear:

$$\varphi(z) \cong 1 - \frac{2\delta}{\zeta}\left(1 - \frac{z}{\zeta}\right) \qquad (15)$$

Eq. (12) simplifies into

$$\frac{dv}{d\varphi} \cong -\tau\, \frac{l\chi_\omega(1)}{\omega\psi_o + p\eta_o}\,(1-\varphi)^\alpha \qquad (16)$$

Here $p(\varphi) = p(1)$, so that $p\eta_o$ is the « critical amplitude » of the shear viscosity near $\varphi = 1$. Similarly, $q(1)\psi_o$ is the critical amplitude of $\psi$. Because we have no idea of the value of $\psi_o$ (this coefficient, conversely to $\eta_o$, is not a characteristic of the interstitial fluid), we may omit the $q(1)$ factor without loss of generality and simply denote $\psi_o$ the critical amplitude of $\psi$. Integrating Eq. (16) gives $v(\varphi,\tau)$. Writing $V_o = v(\varphi_o,\tau)$ and expressing $\tau$ as a function of $\varphi_o$ (Eq. (14)), yields the relation between the shear stress and the boundary velocity:

$$(\omega\psi_o + p\eta_o)\frac{V_o}{\delta} = c\,\tau\left[\frac{\lambda\delta}{\tau\chi_\omega(1)}\right]^{\frac{\alpha-1}{2}} \qquad (17)$$

with $c = \sqrt{2^{(\alpha+1)}/(\alpha+1)}$. In the asymptotic limit ($\varphi_o \cong 1$), the normal force $N$ derived from Eq.(11) scales linearly with the shear stress:

$$N = \frac{\psi_o \tau}{\omega \psi_o + p \eta_o} = \psi_o \frac{V_o}{\delta} \qquad (18)$$

In the same way, combining Eqs (13) and (16) gives the relation between $\zeta$ and $V_o$:

$$\zeta^{3-\alpha} \cong (\alpha+1) \chi_\omega(1)(\omega\psi_o + p\eta_o) \frac{1}{\lambda (2\delta)^{\alpha-1}} V_o \qquad (19)$$

We thus find that the thickness of the suspended layer, in the 2-phase regime (Fig.9d) and in the $\zeta \gg \delta$ limit, varies as a function of $V_o$ according to a power-law of exponent $1/(3-\alpha)$. We end this paragraph with the asymptotic form of the velocity profile, which is a simple power-law too:

$$v(z) \cong V_o \left(1 - \frac{z}{\zeta}\right)^{\alpha+1} \qquad (20)$$

Eq.(20) holds inside the sheared zone ($z \leq \varsigma$). The velocity $\equiv 0$ for $z \geq \varsigma$.

## 5. DISCUSSION

We now discuss the adequacy of the above model to interpret the experimental observations. We suppose that the parallel-disk geometry is locally equivalent to the theoretical infinite-plane geometry, taking $V_o$ as the boundary velocity along the top plate. Recall that the slip-ratio $\mu = V_o/V$ is everywhere $<1$ and only depends on $V$, i.e. $\mu$ is a function of $\Omega r$. Essentially we want to find values for the $\alpha$ exponent such that the computed flow profiles are representative of the experimental ones. The analysis mainly relies on the asymptotic behaviors that we arrived at in the former Section. In principle the equations are valid only when $\zeta \gg \delta$. As we confirm below, this condition is satisfied in our experimental conditions.

*5.1 Velocimetry*

We firstly analyze the velocimetry data, and consider the dependence of $\zeta$ on $V_o$. The data in Fig.6b is best fitted to by Eq.(19) if $\alpha = 1$, giving $\zeta \propto V_o^{0.5}$, not far from the experimental result. Note that this conclusion is unambiguous, since supposing $\alpha = 2$ (or more) is definitely ruled out by the data. We thus find that the viscometric coefficients, $\eta$ and $\psi$, diverge as $(1-\varphi)^{-1}$ near compaction.

An immediate consequence of this result is that the flow profiles, according to Eq.(20), should be parabolic. This statement is tested in Fig.10, where measured values of $v(z,r)/V_o$ are plotted versus $(1-z/\zeta)^2$. The rigth plot corresponds to fixed $\Omega$ and variable $r$, and reversely in the left plot. The conclusion is approximately verified, though the plots show a slight systematic undulation. The deviation is most probably caused by the finite size of the particles, which is not accounted for by the continuous model. Near the top plate, the profile is rounded off roughly on the scale of a single particle ($2a \approx 0.2$ mm), which is enough to cause a visible undulation on the scaled velocity profile.

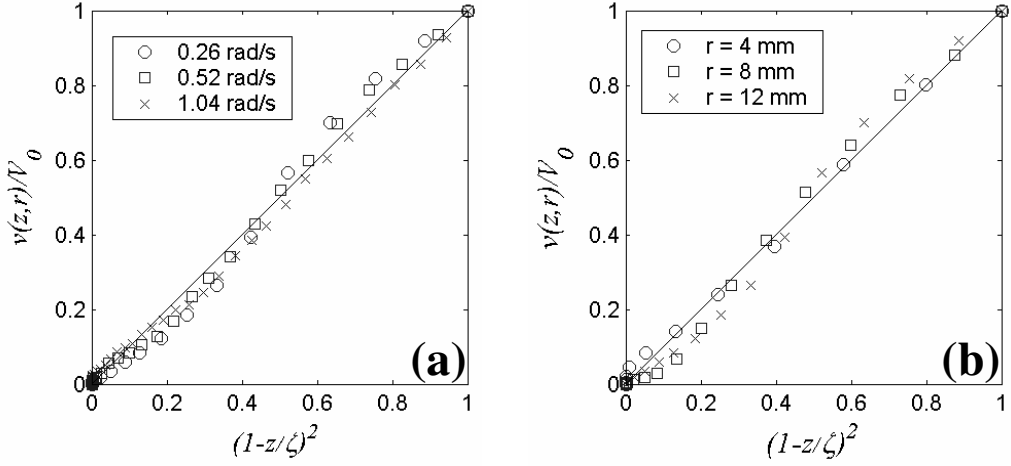

Fig.10: Scaled velocity fields. a : data recorded at constant $r$ = 12mm and variable $\Omega$; b : constant $\Omega$ = 0.26 rad/s and variable $r$.

Note that when $\alpha = 1$ in Eq.(19), the $\delta$ dependence of $\zeta$ is eliminated, i.e. the extent of the shear zone only depends on the boundary velocity, not on the average concentration between the plates. This point turns out important when discussing the applicability of the model to the experimental situation: as we mentioned in Section 2, a small fraction of the suspension under shear escapes out of the parallel-disk space and climbs up in the annular region of the Couette device. Strictly speaking, mass is not conserved in the experimental parallel-plate geometry, contrary to what is supposed in the theory. In other words, $\delta$ is not strictly a constant, and slightly increases with increasing rotor speed. What the above remark implies is that this complication is unimportant, because it does not modify the relation between $\zeta$ and $V_o$. Consequently, our conclusion about the value of $\alpha$ remains valid, in spite of the problem of mass conservation.

*5.2 Rheometry*

To interpret the data from the rheometry experiments, we must find expressions for the total torque and axial force acting on the top plate, or, equivalently, for $\bar{\tau}^*$ and $\bar{\tau}_N$, the reduced stresses defined in paragraph 2.4. We simply integrate the theoretical local torque ($\tau r$) and the local normal force $N$ on the surface of the top disk:

$$\bar{\tau}^* = \frac{2}{\pi R_1^3} \int_0^{R_1} \tau(\Omega r)\, r\, 2\pi r\, dr \tag{21a}$$

$$\bar{\tau}_N = \frac{2}{\pi R_1^2} \int_0^{R_1} N(\Omega r)\, 2\pi r\, dr \tag{21b}$$

Strictly speaking, working out the expression for $\bar{\tau}_N$ needs handling not one but two normal force components[16], and, correspondingly two normal force coefficients, usually denoted $\psi_1$ and $\psi_2$. In practice, the experimental information available to us only provides information on a single coefficient, which we will naively take $=\psi$, the normal stress coefficient defined in the context of the infinite-plate geometry (the discussion on the separate first and second normal force differences will be the matter of a forthcoming paper). Injecting Eqs.(17-18) into Eq.(21) with $\alpha=1$ yields:

$$\bar{\tau}^* = \frac{4\Omega}{R_1^3 \delta}(\omega\psi_o + p\eta_o)\int_0^{R_1}\mu\, r^3\, dr \quad \text{and} \quad \bar{\tau}_N = \frac{4\psi_o \Omega}{R_1^2 \delta}\int_0^{R_1}\mu\, r^2\, dr \qquad (22)$$

In the asymptotic regime (large velocity), the slip ratio saturates to a constant value ($\mu \to \mu_\infty$), resulting in simple rheological laws:

$$\bar{\tau}^* \approx (\omega\psi_o + p\eta_o)\frac{\mu_\infty \Omega R_1}{\delta} \quad \text{and} \quad \bar{\tau}_N \approx \frac{4\psi_o}{3}\frac{\mu_\infty \Omega R_1}{\delta} \qquad (23)$$

$\bar{\tau}^*$ or $\bar{\tau}_N$ are simply linear in $\Omega$ in the $\Omega \to \infty$ limit. At low rotor speed, wall-slip is considerable, especially near the rotation axis where particles hardly move: in this regime, $\mu \propto \Omega r$, and $\bar{\tau}^* \propto \bar{\tau}_N \propto \Omega^2$, correspondingly. This is in line with the experimental record (Fig7a, b). The crossover between both regimes occurs for $\Omega \approx 0.2$ rad/s, i.e. about 4 mm/s near the periphery of the rotor, in good correspondence with the crossover velocity deduced from the wall-slip data (Fig.4c).

*5.3 Parameter values*

The unkown parameters of the model are $p(1)$, $\psi_0$, $\delta$ and $\omega$. Moreover, note that the full $p(\varphi)$ and $q(\varphi)$ functions, whose precise shapes are not known, enter Eqs. (17) and (19) through the $\chi_\omega(1)$ coefficient. For an order-of-magnitude estimate of $p(1)$, $\psi_0$, $\delta$ and $\omega$, we may simply put $p(\varphi) = p$, $q = 1$. This approximation gives $\chi_\omega(1) = 3p\eta_o\psi_o/(\omega\psi_o + p\eta_o)^2$ and simplifies Eq.(19) into: $\zeta^2 \cong 6 A (V_0/\lambda)$, with $A = 6 p\eta_o/(\omega + p\eta_o/\psi_o)$. Fitting the latter equation to the data (Fig.6b) yields the value of $A$, which is found $\cong 55$ poise. Besides, in the high rotor speed regime, the dimensionless quantity $\bar{\tau}^*/\bar{\tau}_N$ takes on a constant value, $B = 3(\omega + p\eta_o/\psi_o)/4$, in agreement with experimental data (Fig.7c). From the experimental slope, $d\bar{\tau}^*/d\bar{\tau}_N \approx 0.11\text{-}0.12$, and combining with $AB = 9p\eta_o/2$, we deduce $p \cong 5.5$ and $\omega + p\eta_o/\psi_o \cong 0.15$. We thus find that $\omega \ll 1$, meaning that the yield-stress effect induced by gravity is much less in the sheared material than at rest. If we simplify this statement into $\omega = 0$, we can further deduce the value of $\psi_0$, which is found $\approx 9.2$ poise.

The conclusion about $\omega$ is not obvious, since the material is jammed by gravity before the onset of shear flow ($\omega \approx 1$ at rest). Apparently, there is no memory of this jamming in the stationary shear flow, in our experimental conditions. This finding is in line with recent propositions about the very meaning of the "yield stress" concept[32,33] : essentially, the yield-stress fluid exists in two states, a jammed one (of infinite viscosity) and a fluid one (of finite viscosity), and the transition between both states is discontinuous, both in stress and shear rate ($\sigma < \sigma_c$, $\dot{v} = 0$ in the jammed state, $\sigma \geq \sigma_c > 0$ and $\dot{v} \geq \dot{v}_c > 0$, in the fluid state). In our description, jamming is simply due to gravity, which creates chains of particles in contact to each other. In this situation, $\omega \approx 1$. We understand the transition to the fluid state ($\omega \cong 0$) as a kind of shear-induced melting of the contact chains. Once melting is achieved, there is no trace left of the initial jamming, leading to our observation that $\omega \ll 1$. Conformity to the phase-transition picture would be complete if a minimum shear-rate ($\dot{v}_c$) were visible in the experimental flow profiles, in the 2-phase regime. This minimum is not detected, but this may be just a matter of accuracy, since the uncertainty on the particle velocity is considerable near the SuSe interface.

Coming back to rheology, if we neglect the $\omega\psi_0$ term, Eq.(17) with $\alpha = 1$ simplifies into $\tau = p\eta_o(V_o/\delta) = [p\eta_o/(1-\overline{\varphi})](V_o/d)$, i.e. the naive form of the shear stress for a Newtonian material of shear viscosity $= p\eta_o/(1-\overline{\varphi})$, in a parallel-plate geometry with a gap $d$. Intuitively we expect this result to be valid in the high shear rate regime, when the material is completely fluidized (Fig.1e). In this limit, the role of gravity becomes negligible, and the material behaves as a suspension of neutrally buoyant particles. What the above analysis arrives at is a non trivial result, i.e. the fact that the viscous stress still obeys the same relation at moderate shear, in the 2-phase regime, where the concentration and local shear rate are far from uniform. Consequently, the rheometry experiment is blind to the structure of the sheared material, as it makes no difference between the 2-phase and 1-phase configurations.

The last parameter, $\delta = d\,(1-\overline{\varphi})$, can be estimated from the slopes of $\overline{\tau}^*(\Omega)$ or $\overline{\tau}_N(\Omega)$ in the high rotor speed limit (linear branches in Fig.7a, b). We find $\delta \cong 390$ μm, in agreement with experimental observations (paragraph 2.2). This result is an "a posteriori" justification for using the asymptotic regime equations of the model: clearly, the condition $\delta/\zeta << 1$ is fulfilled in the experimental domain.

The above value of $\delta$ gives $\overline{\varphi} \cong 0.92$, and with $\overline{\Phi} = 59$ %, the known value of the average solid concentration in the sample, yields $\Phi_m = \overline{\Phi}\,\overline{\varphi} \cong 64$ %. Here $\Phi_m$ must be understood as the critical concentration where the viscometric coefficients diverge (Eqs.(5, 6)). Remarkably, the above estimated value of $\Phi_m$ is close to the random-close-packing concentration ($\cong 64$ %)[10].

*5.4 The $\alpha = 1$ singularity*

We now comment on our essential result, i.e. the $(1-\varphi)^{-1}$ singularity of the viscometric coefficients. A detailed analysis of how the shear viscosity of a concentrated suspension diverges near compaction was worked out by Brady and Morris[19]. These authors made a distinction between the low shear rate and the high shear rate regimes, according to whether the Peclet number is small or large. This number is a measure of the importance of advection relatively to thermal diffusion, and may be defined as $Pe = \dot{v}\,a^2/D$, where $a$ is the particle radius and $D$ the particle diffusion coefficient. In the low shear regime, it is found that the Brownian noise drives a $(1-\varphi)^{-2}$ singularity, i.e. $\alpha = 2$. In the high shear regime, the Brownian noise becomes negligible in the $Pe \to \infty$ limit, and the form of the singularity mainly depends on the existence of finite-range inter-particle repulsive forces. When such forces exist, they prevent the particles from coming in contact to each other, and this still drives a $(1-\varphi)^{-2}$ singularity. Conversely, if hard sphere contact is allowed, a "bare" $(1-\varphi)^{-1}$ singularity is expected.

The Peclet number may be estimated as $Pe \approx \eta_0 a^3 V/d\,kT$, where $kT$ is the thermal energy, and is at least on the order of $10^7$ in our experimental situation. Moreover, nothing prevents solid-solid contacts between the particles. Thus, we are in the simple case where Brownian noise is negligible, and where only hydrodynamic and contact forces are at work. In this context, our statement that $\alpha = 1$ turns out in line with the theoretical prediction for the shear viscosity[18,19]. However it is unclear to us what the theory predicts for the normal force coefficient $\psi$. At this point, it is important to recall our supposition that $\eta$ and $\psi$ would diverge in the same way near compaction ($\alpha = \alpha'$). The proportionality between the average shear stress $\bar{\tau}^*(\Omega)$ and the average normal stress $\bar{\tau}_N(\Omega)$ in the presence of flow localization strongly supports such an assumption.

*5.5 "Shear-induced migration"*
We end this Section with a comment on the so-called "shear-induced migration": this phenomenon, basically a non uniformity in the particle concentration field, has been observed with *density-matched* (neutrally buoyant) particles in situations where the shear rate is not spatially uniform[34-40]. This is so for instance in the parallel-disk geometry, since the plate velocity explicitly depends on the distance to axis ($V = \Omega r$). Migration was explained as being due to competing gradients of particle concentration, shear rate and suspension viscosity, and, in later versions, with the added contribution of gradients of flow lines curvature. Afterwards, it was shown that migration could equivalently be explained with the normal force balance theory[8]. We insist that the non uniformity of particle concentration and the correlated non linearity of the flow in our situation are due the gravity; in other words, setting $\lambda = 0$ in our model leads to a spatially uniform concentration and to a linear flow profile, i.e. no migration. Nevertheless it is well possible that migration in the above defined sense (i.e. independent of gravity) be present in our experiments, for instance as a variation of the average concentration $\bar{\Phi}$ with $r$. This point was not explored yet, but could be in the future with an improved version of the concentration measurement procedure.

**6. CONCLUSION**
In this paper we studied the flow of a very concentrated "slurry", made of negatively buoyant non-colloidal spheres in a viscous liquid. The material was sheared between parallel horizontal disks and observed by optical means, to obtain flow profiles and the related viscous torque and axial force responses. The main goal of the paper was to point out the effects of *gravity* on the observed main features of the shear flow, i.e. the 2-phase structure (fluid-solid), the wall-slip phenomenon, and the non-linearity of the flow.

To interpret the experimental data, we proposed a "mean-field" model for the ideal shear geometry between two horizontal infinite planes and based on the principles of the normal force balance theory. We studied the adequacy of the model to describe the shear flow inside the real parallel-disk device, up to a limit distance from the axis beyond which the flow becomes sensitive to the finite size of the device.

We found that the model correctly accounts for the observed trends, assuming a unique $(\Phi_m - \Phi)^{-1}$ singularity for both viscometric coefficients. A rather unexpected conclusion is the adequacy of the theoretical description, which is based on simple local and mean-field equations, for the very high concentrations of interest (up to 92% of $\Phi_m$). This is probably due to the freedom left to the material to slightly dilate using the available space around the parallel-disk zone. We recently carried out experiments with no excess oil, which amounts to forbidding the overall dilatancy of the material. In this case, the local equivalence to the infinite plate geometry is lost, as varying $\Omega$ or $r$ turn out non equivalent.

We end this Section and the article by tentatively listing what we believe to be important unanswered questions, and making suggestions, correspondingly:

- We were not able to obtain reliable concentration profiles in the shear flow from the experiments. This information would be valuable, since the theory makes a simple prediction that could be tested (Eq. (15)).
- The wall-slip phenomenon, which is an important feature of the material's response to shear, lacks a theoretical description. This problem is not restricted to granular materials (either dry or wet), as it was already noticed with polymer solutions, emulsions[41] and even molecular fluids on the nanometer scale[25]. Based on the experimental observations, our intuition is that wall-slip and the fluctuations of the particles velocity are related effects. Both phenomena could possibly be accounted for by a dynamical version (time-dependent) of the model, inspired from the theory of Picard et al.[33] for the shear flow of yield stress fluids. Indeed these authors predicted the existence of flow oscillations associated with shear-banding in situations where the average shear rate is imposed, as is the case in our experiments.

**ACKNOWLEDGEMENTS**

This work is financially supported by DGA, SNPE-SME and Région-Aquitaine. We thank C. Barentin, G. Ovarlez, P. Mills, S. Marchetto, M. Gaudré and C. Marraud for fruitful discussions, and the CRPP Instrumentation group for technical support.

**APPENDIX A**

We consider a homogeneous slurry in a plane Couette geometry (Fig.2b). The slurry, of average viscosity $\bar{\eta}$, fills the space between parallel plates (gap $d$) and the pure oil (viscosity $\eta_o$) partially fills the annular volume between cylinders (height $h$). For negligible inertial effects, and to a first approximation, the viscous torque $\Gamma$ may be decomposed into three contributions, arising from the bottom surface ($\Sigma^*$) and from the peripheral areas ($\Sigma_a, \Sigma_b$) of the outer cylinder (Fig.A1) :

$$\Gamma \approx \int_0^{R_1} \tau^* r \, 2\pi r \, dr + \int_0^d \tau_a R_2 \, 2\pi r \, dz + \int_0^h \tau_b R_2 \, 2\pi r \, dz \qquad (A1)$$

with $\tau^* \approx \bar{\eta}\, \Omega\, r/d$, $\tau_a \approx \bar{\eta}\, \Omega\, R_2/s$, $\tau_b \approx \eta_o\, \Omega\, R_2/e$, where $e = R_2 - R_1$, and $s = (e^2 + z^2)^{1/2}$. Eq.(A1) leads to an approximate expression for the average shear stress $\bar{\tau} = \Gamma/(\pi R_1^3/2)$:

$$\bar{\tau} \approx 4\,\Omega \left[ \bar{\eta}\, \frac{R_1}{4d} + \bar{\eta}\, \frac{R_2^3}{R_1^3} Ln\left[ \left((d/e)^2 + 1\right)^{1/2} + d/e \right] + \eta_o\, \frac{R_2^3}{R_1^3}\, \frac{h}{e} \right] \qquad (A2)$$

If $R_2 \cong R_1$ and $R_2 - R_1 << d$, as is the case in our experiments, the average shear stress $\bar{\tau}^* = \Gamma^*/(\pi R_1^3/2)$ corresponding to the parallel-plate geometry is given by:

$$\bar{\tau}^* \approx \bar{\eta}\frac{\Omega R_1}{d} \approx \frac{\bar{\tau} - 4\eta_o \Omega \frac{h}{R_2 - R_1}}{1 + \frac{4d}{R_2} Ln\left[2d/(R_2 - R_1)\right]} \quad (A3)$$

We tested the validity of Eq.(A.2) with samples constituted of pure oil more or less filling the gap between cylinders, i.e. when $h$ is varied (Fig.2b). The graphs below (Fig.A1) show that Eq.(A3) correctly predicts the increase of $\bar{\tau}$ when $h$ is increased. We supposed that the same approximation holds in general and used Eq.(A3) to estimate the corrected average shear stress $\bar{\tau}^*$ for the slurry.

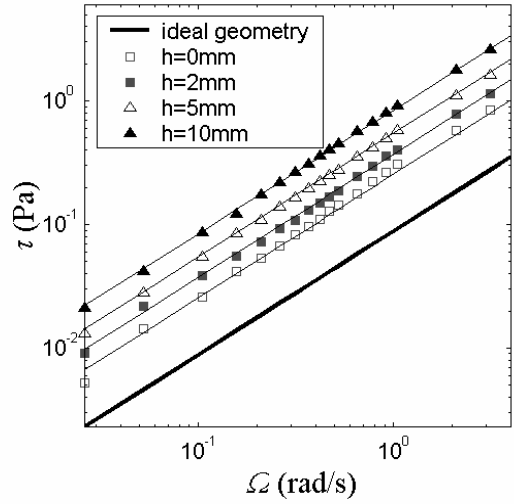

Fig.A1: Average shear stress $\bar{\tau}$ versus rotor velocity $\Omega$ for pure oil filling the space between parallel plates and for $h$ = 0, 2, 5, 10 mm. The corresponding straight lines are the predictions from Eq.(A3) with $\bar{\eta} = \eta_o \cong 0.25$ Poise and $d$ = 5 mm. The bold solid line indicates the $\bar{\tau}^*(\Omega)$ dependence for the parallel disk geometry without edge effects.

## APPENDIX B

A numerical scheme was developed to solve the coupled Eqs.(2, 3) for viscometric coefficients given by Eqs.(5, 6) with $\alpha' = \alpha$, $p(\varphi) = p$, $q(\varphi) = 1$. The concentration profile $\varphi(\bar{z},\tau)$ and the extent $\zeta(\bar{z},\tau)$ of the flowing region were derived from $\bar{z}(\varphi) = X_\omega(\varphi) + \bar{z}_1$ and Eq.(4) for mass conservation through the graphical procedure described in paragraph 4.2. The average shear stress $\bar{\tau}*(\Omega)$ and the average normal stress $\bar{\tau}_N(\Omega)$ were calculated from the integrals (21a) and (21b) with an expression for the slip ratio $\mu(V)$ obtained from a fit to experimental data (Fig.6b). In the context of experimental conditions ($\bar{\Phi} = 0.59$ and $\Phi_m = 0.64$), the full numerical model gives concentration and velocity profiles in close agreement with asymptotic predictions from Eqs.(15-19, 20) as shown in Fig.B1 below.

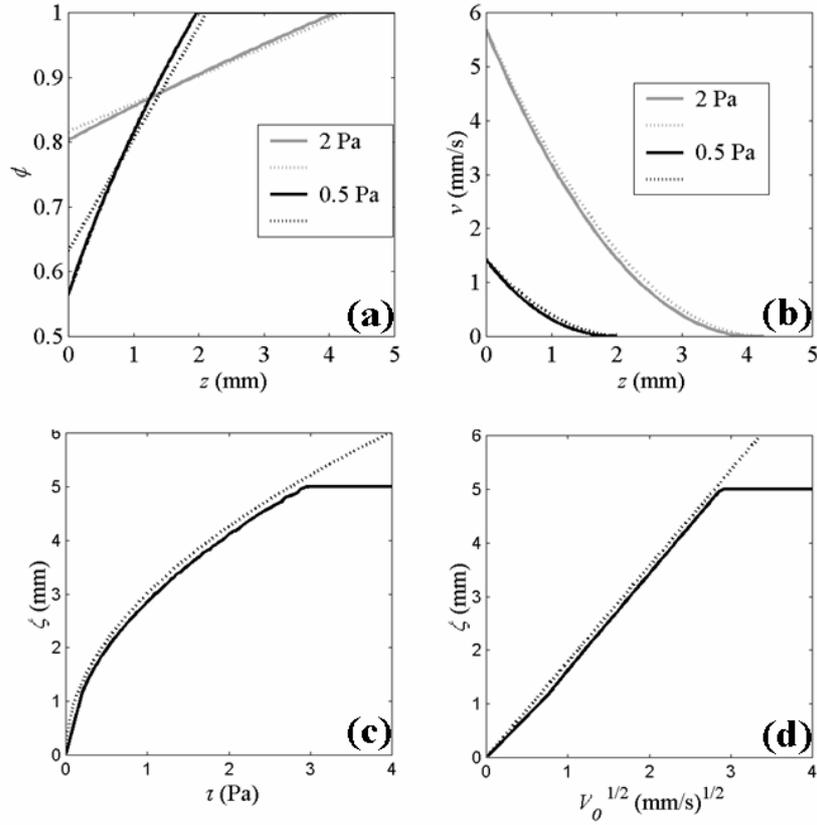

Fig.B1: Theoretical concentration profiles (a), velocity profiles (b) and thickness $\zeta$ of the flowing region versus the tangential stress (c) and versus the slurry boundary velocity (d), in infinite-plate geometry. Solid and dotted lines correspond to the full numerical solution and to asymptotic expressions Eqs.(15-19, 20), respectively, with $\omega$ =0, $p$ =5.5, $\eta_o$ = 0.25 Poise, $\psi_o$ = 9.2 Poise, $R_1$ = 18.5mm, $d$ = 5 mm, $\overline{\Phi} = 0.59$ and $\Phi_m = 0.64$.

# REFERENCES

[1] D. Leighton and A. Acrivos, "Viscous resuspension," Chem. Engng Sci. **41**, 1377 (1986).

[2] U. Schaflinger, A. Acrivos, and B. Zhang, "Viscous resuspension of a sediment within a laminar and stratified flow," Int. J. Multiphase Flow **16**, 567 (1990).

[3] U. Schaflinger, A. Acrivos and H. Stibi, "An experimental study of viscous resuspension in a pressure-driven plane channel flow," Int. J. Multiphase Flow **21**, 693 (1995).

[4] J. F. Morris and J. F. Brady, "Pressure driven flow of a suspension: Buoyancy effects," Int. J. Multiphase Flow **24**, 105 (1998).

[5] T. Loimer and U. Schaflinger, "The effect of interfacial instabilities in a stratified resuspension flow on the pressure loss," Phys. Fluids **10**, 2737 (1998).

[6] F. Charru and H. Mouilleron-Arnould, "Instability of a bed of particles sheared by a viscous flow," J. Fluid Mech. **452**, 303 (2002).

[7] P. R. Nott and J. F. Brady, "Pressure-driven flow of suspensions: simulation and theory," J. Fluid Mech. **275**, 157 (1994).

[8] J. F. Morris and F. Boulay, "Curvilinear flows of noncolloidal suspensions: The role of normal stresses," J. Rheol. **43**, 1213 (1999).

[9] I. E. Zarraga, D. A. Hill and D. T. Leighton, "The characterization of the total stress of concentrated suspensions of noncolloidal spheres in Newtonian fluids," J. Rheol. **44**, 185 (2000).

[10] G. Y. Onoda and E. G. Liniger, "Random loose packings of uniform spheres and the dilatancy onset," Phys. Rev. Lett. **64**, 2727 (1990).

[11] C. Barentin, E. Azanza and B. Pouligny, "Flow and segregation in sheared granular slurries," Europhys. Lett. **66**, 139 (2004).

[12] O. Pouliquen, "Scaling laws in granular flows down rough inclined planes," Phys. Fluids 11, 542 (1999).

[13] R. J. Adrian, "Particle-imaging technique for experimental fluid mechanics," Ann. Rev. Fluid Mech. **23**, 261 (1991).

[14] N. Tetlow, A. L. Graham, M. S. Ingber, S. R. Subia, L. A. Mondy and S. A. Altobelli, "Particle migration in a Couette apparatus: experiment and modelling," J. Rheol. **42**, 307 (1998).

[15] W. Losert, L. Bocquet, T. C. Lubensky and J. P. Gollub, "Particle dynamics in sheared granular matter," Phys. Rev. Lett. **85**, 1428 (2000).

[16] R. B. Bird, R. C. Armstrong and O. Hassager, "Dynamics of polymeric liquids," vol. 1 *Fluid Mechanics* (John Wiley and Sons, New York, 1987).

[17] F. Gadala-Maria and A. Acrivos, "Shear induced structure in a concentrated suspension of solid spheres," J. Rheol. **24**, 799 (1980).

[18] T. N. Phung, J. F. Brady and G. Bossis, "Stokesian Dynamics simulation of Brownian suspensions," J. Fluid Mech. **313**, 181 (1996).

[19] J. F. Brady and J. F. Morris, "Microstructure of strongly sheared suspensions and its impact on rheology and diffusion," J. Fluid Mech. **348**, 103 (1997).

[20] GDR MiDi, "On dense granular flows," Eur. Phys. J. E **14**, 341-365 (2004).

[21] G. Josserand, P. Y. Lagrée and D. Lhuillier, "Stationary shear flows of dense granular materials : a tentative continuum modelling, " Eur. Phys. J. 14, 127-135 (2004).

[22] P. Mills and P. Snabre, "Rheology and structure of concentrated suspensions of hard spheres. Shear-induced particle migration," J. Phys. II France **5**, 1597 (1995).

[23] P. G. de Gennes, "Ecoulements viscométriques de polymères enchevêtrés, " C. R. Acad. Sci. Paris B **288**, 219 (1979).

[24] E. Lauga and H. A. Stone, "Effective slip in pressure-driven Stokes flow," J. Fluid Mech. 489, 55 (2003).

[25] C. Cottin-Bizonne, C. Barentin, E. Charlaix, L. Bocquet and J.-L. Barrat, "Dynamics of simple liquids at heterogeneous surfaces: molecular dynamics simulations and hydrodynamic description," Eur. Phys. J. E **15**, 427-438 (2004).

[26] N. A. Frankel and A. Acrivos, "On the viscosity of a concentrated suspension of solid spheres," Chem. Engng. Sci. **22**, 847 (1967).

[27] K. C. Nunan and J. B. Keller, "Effective viscosity of a periodic suspension," J. Fluid Mech. **142**, 269 (1984).

[28] J. F. Brady, "The rheological behaviour of concentrated colloidal dispersions," J. Chem. Phys. **99**, 567 (1993).

[29] J. F. Brady, "The long-time self-diffusivity in concentrated colloidal dispersions," J. Fluid Mech. **272**, 109 (1994).

[30] P. Mills and P. Snabre, The fractal concept in the rheology of concentrated suspensions,. Rheol. Acta **26**, 105 (1988).

[31] I. M. Krieger, "Rheology of monodisperse lattices," Adv. Colloid Interface Sci. **2**, 111 (1972).

[32] P. Coussot, Q.D. Nguyen, H.T. Huynh and D. Bonn, "Avalanche behavior in yield stress fluids," Phys. Rev. Lett. **88**, 175501 (2002).

[33] G. Picard, A. Ajdari, L. Bocquet and F. Lequeux, "Simple model for heterogeneous flows of yield stess fluids," Phys. Rev. E **66**, 051501 (2002).

[34] D.Leighton and A.Acrivos, "The shear-induced migration of particles in concentrated suspensions," J. Fluid Mech. **181**, 415 (1987).

[35] J. R. Abbott, N. Tetlow, A. L. Graham, S. A. Altobelli, Eiichi Fukushima, L. A. Mondy and T. S. Stephens, "Experimental observations of particle migration in concentrated suspensions: Couette flow," J. Rheol. **35**, 773 (1991).

[36] R. J. Philips, R. C. Armstrong, R. A. Brown, A. L. Graham and J. R. Abbott, "A constitutive equation for concentrated suspensions that accounts for shear-induced particle migration," Phys. Fluids A **4**, 30 (1992).

[37] P. Mills and P. Snabre, "Rheology and structure of concentrated suspensions of hard spheres. Shear-induced particle migration," J. Phys. (France) II **5**, 1597 (1995).

[38] R. E. Hampton, A. A. Mammoli, A. L. Graham, N. Tetlow and S. A. Altobelli, "Migration of particles undergoing pressure-driven flow in a circular conduit," J. Rheol. **41**, 621 (1997).

[39] G. P. Krishnan, S. Beimfohr and D. Leighton, "Shear induced radial segregation in bidisperse suspensions," J. Fluid Mech. **321**, 371 (1996).

[40] S. R. Subia, M. S. Ingber, L. A. Mondy, S. A. Altobelli and A. L. Graham, "Modelling of concentrated suspensions using a continuum constitutive equation," J. Fluid Mech. **373**, 193 (1998).

[41]J. B. Salmon, L. Bécu, S. Manneville and A. Colin, "Towards local rheology of emulsions under Couette flow using Dynamic Light Scattering," Eur. Phys. J. E 10, 209 (2003).